\documentclass{revtex4}

\usepackage{graphicx}
\usepackage{amssymb}
\usepackage{amsmath}

\begin{document}

\title{Cosmological dynamics of scalar field with non-minimal kinetic term}

\author{
H.~Kr\"{o}ger$^{a}\footnote{Email: hkroger@phy.ulaval.ca}$,
G.~Melkonyan$^{a}\footnote{Email:
gmelkony@phy.ulaval.ca}$,
S.G.~Rubin$^{b,c}\footnote{Email:
sergeirubin@mtu-net.ru}$}

\affiliation{ $^{a}$ {\small\sl D\'{e}partement de Physique, Universit\'{e}
Laval, Qu\'{e}bec} \\ $^{b}$ {\small\sl Moscow State Engineering Physics
Institute, Moscow, Russia}
\\ $^{c}$ {\small\sl Center for Cosmoparticle Physics "Cosmion", Moscow, Russia} }

\begin{abstract}
{We investigate dynamics of scalar field with non-minimal kinetic term.
Nontrivial behavior of the field in the vicinity of singular points of kinetic
term is observed. In particular, the singular points could serve as attractor
for classical solutions.}
\end{abstract}

\maketitle

key words: kinetic term, scalar, gravity, inflation, vacuum energy

\section{Introduction}
In this paper we investigate dynamics of scalar field theory with the kinetic
term dependent on the field value. This hypothesis provides an additional
resources in an explanation of observational data. The small value of
cosmological term, consistent with recent experimental data \cite{Knop03} can
be explained using a non-canonical form of the kinetic term in the scalar field
(like in the quintessential model \cite{Damour99,Chiba99}). Substantial review
of different ways of origin of the cosmological term can be found in
\cite{Sahni00}. A non-trivial kinetic term could be responsible for a new
coupling between adiabatic and entropy perturbations \cite{Marco02}.

Such a model may be considered as the restricted scalar-tensor theory (STT) of
gravity \cite{Bergmann}. There is much current interest in the STT, because it
allows for explanations of variety of cosmological phenomena
\cite{Boisseau00,Wang,Esp03} like, e.g., inflation at a low energy scale
\cite{Morris01} and dark matter caused by phantom fields \cite{Singh03}. This
theory, being a generalization of general relativity, possesses a much richer
structure then the latter. The model has a sector of a scalar field and it
seems natural to relate this component to the inflation. Historically, the
first internally self-consistent inflationary model \cite{Star79} appears to be
equivalent to the special limiting case of STT \cite{Equiv}.

The exact form of STT action has not been derived yet from a more fundamental
theory. Thus a large number of phenomenological potentials and kinetic terms
are considered in current models. Some of those can be excluded being
inconsistent with astrophysical data. Still a substantial number of them
remains topical. In this connection two questions can be raised and need to be
answered: (i) What shape of potential is realized in our universe? Finding an
answer would require that the theoretical prediction from such model potential
would fit most observational data. Suppose one would find answer, then the next
question is: (ii) Why is this particular shape of potential (or kinetic term)
realized in nature? What are the underlying reasons?

Some theoretical hints on the form of the potential have been given by
supergravity, which  predicts an infinite power series expansion in the scalar
field potential \cite{Nilles}. Its minima, if they exist, correspond to the
stationary states of the field. In the low energy regime it is reasonable to
retain only a few terms (lowest powers in the Taylor expansion) of the scalar
field \cite{Lyth96}. However, the potential caused, e.g., by supergravity could
correspond to a function with infinite set of the potential minima. There is
still no physical law which limits the potential to possess only a finite
number of minima. In the vicinity of each of the minima the potential has an
individual form and hence the universe associated with such minimum may be
individually different from others. Our own universe is associated with a
particular potential minimum, not necessarily located at $\varphi = 0$. The
supposition that the potential possesses infinite number of randomly
distributed minima is self-consistent \cite{Ru42a}. A similar behavior may hold
also for the kinetic term of STT.

The observed smallness of the value of the $\Lambda$ term  is explained usually
on the basis of a more fundamental theory like supergravity or the anthropic
principle. Our point of view is that we have to merge these approaches. The
more fundamental theory supplies us with an infinite set of minima of the
potential. These minima having an individual shape are responsible for the
formation of those universes used in the anthropic picture.

Actually, any way of introducing a scalar field leads to a theory similar to
STT provided that quantum corrections are taken into account. Indeed, STT
arises necessarily in ordinary quantum field theory with a scalar field. For
example, it is a standard result that the kinetic term of the effective action
acquires a multiplier which is a function of the field \cite{Itzykson}.

Here we consider an action with a non-trivial kinetic term of the following
form
\begin{eqnarray}\label{one}
S=\int d^4 x \sqrt{-g} \left[ \frac{R}{16\pi G} + \frac12
K(\varphi )\partial _\mu \varphi \partial ^\mu \varphi -
V(\varphi) \right] ~ .
\end{eqnarray}
In this paper we concentrate on the influence of singular points of the kinetic
term on the scalar field dynamics. The singular points are not exceptional
cases. E.g., the well known Brans - Dicke model \cite{Brans} does contain a
singularity at zero value of the field $\varphi$. If a multiplier of the Ricci
scalar equals to zero at some point in the Jordan frame, the kinetic term will
be singular in the Einstein frame (see e.g. \cite{Morris01}). Quintessential
models with negative power law \cite{Peebles} are another well known example
for the potential of such a sort.

By a suitable change of variable the function $K$ can be transformed to become
$K(\varphi )=\pm 1$. Such behavior is physical during some inflationary period
\cite{Ru} but not at some recent epoch where the field fluctuates around the
singular point. The similar problem is discussed in refs. \cite{Star81},
\cite{Bronnikov02} and \cite{Esp} in the framework of more general case of STT.

The general form of the kinetic term considered in this paper is
\begin{eqnarray}\label{two}
K(\varphi )= M^n /(\varphi - \varphi _s )^n .
\end{eqnarray}
This form is correct at least in the vicinity of the singularity if $n>0$, or
zero if $n<0$. Here $M$ is some parameter (its value will be discussed below).

The equation of motion has the form
\begin{eqnarray}\label{threea}
K(\varphi)[ \ddot{\varphi} + 3H\dot{\varphi} ] + \frac12
K(\varphi)^\prime \dot{\varphi}^2 + V(\varphi )^\prime =0
\end{eqnarray}
in the Friedmann-Robertson-Walker universe where $H$ denotes the
Hubble parameter. Keeping in mind expression (\ref{two}) we have
\begin{eqnarray}\label{threeb}
\ddot{\varphi} + 3H\dot{\varphi} - \frac{n}{2(\varphi - \varphi _s
)} \dot{\varphi}^2 + V(\varphi _s )^\prime (\varphi - \varphi _s)^n
/M^n =0.
\end{eqnarray}

We observe that  $\varphi=\varphi _s$ is a stationary solution for any smooth
potential $V$ and $n>0$. The cosmological energy density of the vacuum is
connected usually with one of the potential minima. In the case considered here
it is not like this - it could be located at the singular point of the kinetic
term $K(\varphi )$.

In the following we limit ourselves by considering the first two
terms in Taylor series of the potential and shift the field
variable so that the potential acquires the simple form
\begin{eqnarray}
\label{potential}
V(\varphi )=V_0 +m^2\varphi ^2 /2.
\end{eqnarray}
To facilitate a more detailed analysis, a new auxiliary variable $\chi$ is
useful. We suggest the substitution of variables $\varphi
\to \chi$ in the following manner
\begin{eqnarray}\label{three}
d\chi =\pm \sqrt{K(\varphi)}d\varphi ,\quad K(\varphi)>0.
\end{eqnarray}
An action in terms of the auxiliary field $\chi$ has the form
\begin{eqnarray}\label{five}
S=\int d^4 x \sqrt{-g}\{ \frac{R}{16\pi G} + sgn (\varphi - \varphi_s )
\frac12\partial _\mu \chi \partial ^\mu \chi - U(\chi) \},
\end{eqnarray}
where the potential $U(\chi)\equiv V(\varphi (\chi ))$ 'partly
discontinues' function. Its form depends on the form of initial
potential $V(\varphi )$, the form of kinetic term and a position of
the singularities.
Now let us consider particular cases of $K(\varphi )$.
\section{The case $n=1$}
\begin{figure}[tbp]
\includegraphics[width=7cm,height=5cm]{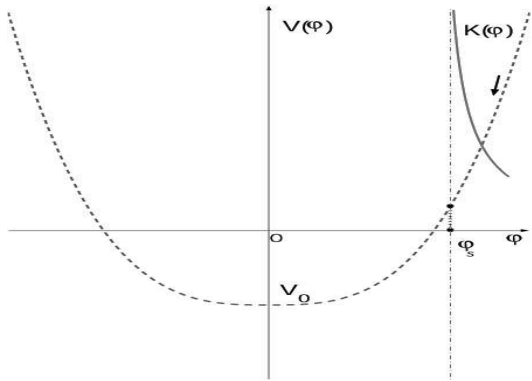}
\caption{The position of the singular point of the kinetic term.} \label{1}
\end{figure}

The kinetic term (\ref{two}) is $K(\varphi )=M/(\varphi -\varphi
_s )$ and Eq.(\ref{three}) reads
\begin{eqnarray}\label{four}
\varphi = \varphi _s  +  sgn (\varphi - \varphi_s )\chi^2 /4M
\end{eqnarray}
with the potential
\begin{eqnarray}\label{fivea}
U(\chi)\equiv V(\varphi (\chi)) = V_0 +\frac{m^2 }{2} (\varphi _s + sgn
(\varphi - \varphi_s )\frac{\chi^2}{4M} )^2 ; \varphi_s >0; |\chi |<\infty .
\end{eqnarray}
The situation is illustrated by Figure 1. If the physical field $\varphi
<\varphi _s $, than the auxiliary field $\chi$ behaves like the phantom field
\cite{Phantom}; if the the field $\varphi$ fluctuates around $\varphi _s $ the
substitution (\ref{three}) is meaningless.

The inflationary scenario on the basis of potential (\ref{fivea}) is well known
and we discuss only its final stage. If  $\varphi _s =0$ and $V_0 =0$ exactly,
we obtain a potential $U\sim \chi ^4$. If $\varphi _s >0$, the auxiliary field
$\chi$ oscillates, according to classical equation, with damping around its
final position at $\chi =0$ - see Figure 2.

\begin{figure}[tbp]
\includegraphics[width=8cm,height=6cm]{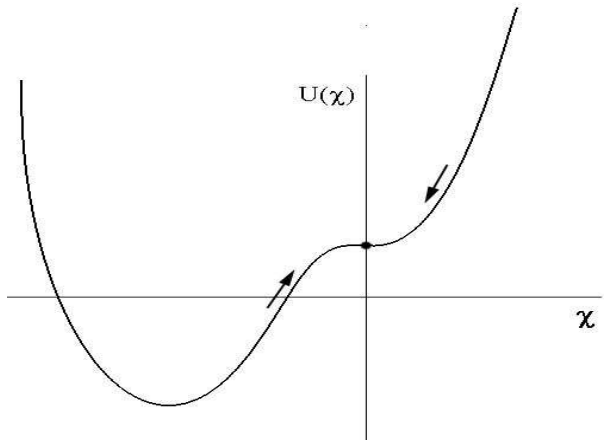}
\caption{The form of the potential for the case $n=1$. Chosen branches are:
$\varphi > \varphi_s \leftrightarrow \chi >0;  \varphi < \varphi_s
\leftrightarrow \chi <0$. In the latter case the auxiliary field $\chi$ behaves
like phantom field moving classically to the local maximum}\label{2}
\end{figure}

For the physical field $\varphi$, the motion looks like oscillations around the
point $\varphi _s $. Finally, the field is captured in the vicinity of the
singular point, supplying an energy density of the vacuum equal to
\begin{eqnarray}\label{VacEn}
\Lambda =V_0 +\frac12 m^2 \varphi _s ^2,
\end{eqnarray}
rather than $\Lambda =V_0$ as it could be expected from expression
(\ref{potential}).

The problem of smallness of the vacuum energy density, $\Lambda
=10^{-123}M_P^4$, is still topical. In other words, what physical reason
adjusts the location of the singular point $\varphi _s$ such that it lies in
the extremely tiny interval $\sqrt{-2V_0 /m^2}\div \sqrt{-2V_0 /m^2+2\Lambda
/m^2}$? In the view of the above discussion, we could expect that some minima
$\varphi _m$ of the potential could incorporate critical point(s) $\varphi _s$
at specific distance from this minimum. Now the problem is reformulated as:
``what part of an infinite amount of the minima contains singular points
located so closely to the minima?'' It seems plausible that this part is very
small, but not zero, due to an infinite number of minima. Only this part is
important - it represents those vacua where galaxies could be formed
\cite{Weinberg}.
\section{The case $n=2$}
In this case, the inflationary period is more interesting than in the previous
case, so we will discuss it more thoroughly. The kinetic term has the form
$K(\varphi)\equiv K_s (\varphi )=M^2 /(\varphi -\varphi_s )^2 $. The new
auxiliary field is connected with the physical one by

$$\chi =  M \ln {\left| \frac{\varphi -\varphi_s}{\varphi
_s}\right|},$$
so that the potential has the form
\begin{eqnarray}\label{pot2}
U(\chi )=\frac12 m^2 \varphi_s^2 (1+ sgn(\varphi _s)\cdot
sgn(\varphi-\varphi _s)\cdot e^{\chi /M })^2 +V_0 .
\end{eqnarray}
If $\varphi _s >0$ and $\varphi >\varphi _s$, the potential mimics the
quintessential model with nonzero vacuum energy density $\Lambda =\frac12 m^2
\varphi_s^2 +V_0$.

The case  $\varphi _s <0$ and $\varphi >\varphi _s$ is much more
interesting. This potential is highly asymmetric so that a behavior
of inflaton is rather different at $\chi < 0 $ and at $\chi > 0 $
\begin{figure}[tbp]
\includegraphics[width=10cm,height=5cm]{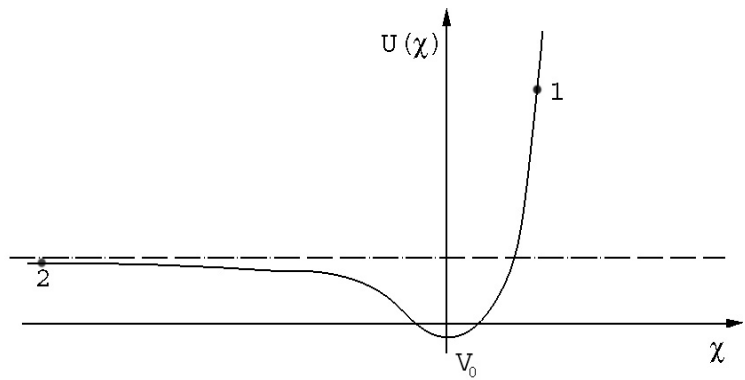}
\caption{The form of the potential of the auxiliary field}\label{4}
\end{figure}
Let us suppose that inflation starts with $\chi_{in} >0$, i.e. point $1$ in
Figure 3. The picture is similar to the improved quintessence potential
\cite{Albrecht}, without problems of the radiation-dominated stage during Big
Bang nucleosynthesis. A modern epoch is characterized by a large negative value
of the field $\chi$. It slowly varies along the flat part of the potential with
exponentially slow variation of the vacuum energy density around the value
(\ref{VacEn}).

It is worth to discuss another possibility, when the inflation starts at $\chi
=\chi _{in}<0$, see Figure 3, point 2. The form of the potential $\chi \to
-\infty$ is rather plain being suitable for slow rolling, so it is reasonable
to expect some new results. Indeed, our estimations indicate that the inflation
could take place at the parameter values $M\sim M_P$ (Planck scale), $m\sim
\varphi_s \sim 10^{-3}M_P $ (GUT scale) and $|\chi |\sim M_p \ln{5}$.
Evidently, this inflationary model does not require fine tuning of its
parameters. The inflation ends at the minimum of the potential, $\chi =0$ and
the final energy density of the vacuum becomes $\Lambda =V_0$.
\section{The case $n=-1$}
A nontrivial situation takes place if the kinetic function has zero value at
some point, i.e. $K(\varphi )=(\varphi -\varphi _s )/M $. Eq.\ref{three} leads
to the classical equation
\begin{eqnarray}\label{threec}
(\varphi_s - \varphi )\cdot (\ddot{\varphi} + 3H\dot{\varphi}) -
\frac{1}{2} \dot{\varphi}^2 + M\cdot V(\varphi )^\prime =0
\end{eqnarray}
This equation is quite uncommon. Indeed, if the zero point $\varphi _s$ of the
kinetic term does not coincide exactly with the position of the potential
minimum at $\varphi =0$, the point $\varphi = \varphi _s$ is not a stationary
solution of this equation. On the other hand, the point $\varphi_s$ is some
kind of attractor. Namely, if the field value is larger than $\varphi_s$, then
$K(\varphi)>0$ and we have standard rolling of the field down to the singular
point $\varphi _s$. If the field value is smaller than $\varphi_s$, then
$K(\varphi)<0$. The field behaves like the phantom field \cite{Phantom},
climbing up to the potential and thus tending toward the point $\varphi_s$.
Classically, the situation looks very strange - the singular point attracts the
solution, but forbids it to stay there forever. Evidently, the field fluctuates
stochastically around the singular point. Additional discussion on this problem
can be found in the paper \cite{Frolov02}.

Let the initial field value $\varphi=\varphi _{in}> \varphi _s$. An appropriate
variable substitution (\ref{three}) looks like
$$\varphi =\varphi_s +sgn (\varphi -\varphi _{s})\cdot \gamma |\chi
|^{2/3},\quad \gamma \equiv (3\sqrt{M}/2)^{2/3} \quad \varphi > \varphi _s .$$
The potential of the auxiliary field $\chi$ becomes
\begin{eqnarray}\label{threed}
U(\chi )=\frac12 m^2 (\varphi_s  +sgn (\varphi -\varphi _{s})\cdot \gamma |\chi
|^{2/3})^2 +V_0 .
\end{eqnarray}
$U(\chi )$ is finite at $\chi =0$ but its derivative is singular,
$$U_{\chi \rightarrow +0}^{\prime }=-\frac{2}{3}m^{2}\varphi _{s}^{2}\gamma
|\chi |^{-1/3}.$$

The potential (\ref{threed}) behaves like $\chi ^{4/3}$ at large field values.
It leads to standard inflation with moderate fine tuning of the parameters. If
$\varphi _s >0$, the field $\varphi$ will oscillate around critical point and
energy density (\ref{VacEn}).
\section{Discussion}
It is shown that the region where the kinetic term changes its sign gives new
possibilities for the scalar field dynamics. It takes place even for the
simplest form of potential. Depending on the position of the singular point of
the kinetic term specific forms of the potential of the auxiliary field can be
obtained. One of the main result is that the scalar field could be located in
the singular points of the kinetic term rather than in minima of the potential.
Another interesting result is that if the kinetic term becomes zero at the
singular point, the behavior of scalar field acquires stochastic character.

The problem of the cosmological $\Lambda$ - term acquires another sense. One
has to explain the extremal proximity of the singular point of the kinetic term
and zero point of the potential. Such situation seems absolutely accidental.
Moreover, the probability of such event to occur is very small. The problem can
be solved in the framework of the random potential \cite{Ru42a} and a kinetic
term of the scalar field discussed in the Introduction. In this case we have
infinite number of singular points of the kinetic term. Fluctuations of the
scalar field which were created at high energies moves classically to
stationary points. Those of them who reach stationary points with appropriate
energy density could form a universe similar to our Universe. This energy
density ($\sim 10^{-123}M_P ^4$) is the result of small value of a concrete
potential minimum or small value of the difference $\varphi_s - \varphi_0$,
$\varphi_0$ is a zero of the potential. The fraction of such universes is
extremely small, but nevertheless is finite because of infinite number of
stationary states.

The abundance of the variants discussed in the paper is based on the simplest
forms of the potential and the kinetic terms. Nevertheless, we reproduced the
quintessence feature, proposed a new possibility for inflation without fine
tuning and discussed a fluctuating state of the scalar field as a final state
which be could realized in the Universe.

One of us (S.G.R) is grateful to K. A. Bronnikov and V. M. Zhuravlev for
discussion. The work was partially performed in the framework of Russian State
contract $40.022.1.1.1106$, RFBR grant $02-02-17490$ and of grant of Russian
Universities UR.02.01.026.

\end{document}